\documentclass[
	aps, prd, reprint,
	10pt, notitlepage,
        floats, floatfix,
	amsmath, amssymb, amsfonts, 
	superscriptaddress,
	showpacs, showkeys,
	nofootinbib,
 	longbibliography,
]{revtex4-1}

\renewcommand\numberwithin

\usepackage{graphicx} 
\usepackage[usenames,dvipsnames]{xcolor}
\usepackage{xspace} 
\usepackage{bm} 
\usepackage{amsmath,amsfonts,amssymb,amsthm}
\usepackage[utf8]{inputenc} 

\xdefinecolor{mylinkcolor}{rgb}{0,0,0.5}
\usepackage[
	bookmarksnumbered, bookmarksopen, bookmarksopenlevel=2,
	breaklinks=true,
	colorlinks=true, filecolor=mylinkcolor, citecolor=mylinkcolor,
	linkcolor=mylinkcolor, urlcolor=mylinkcolor, menucolor=mylinkcolor,
]{hyperref}
\usepackage{verbatim}
\usepackage{mathrsfs}
\usepackage{color}
\usepackage{enumitem}
\usepackage[dvipsnames]{xcolor}
\usepackage{comment}

\newcommand{\nin}{\noindent}
\newcommand{\nnm}{\nonumber}
\newcommand{\doe}{\partial}
\newcommand{\be}{\begin{equation}}
\newcommand{\ee}{\end{equation}}
\newcommand{\bse}{\begin{subequations}}
\newcommand{\ese}{\end{subequations}}

\newcommand{\mr}{\mathrm}
\newcommand{\tr}{\textrm}
\newcommand{\mc}{\mathcal}
\newcommand{\py}{\phantom{yo}}
\newcommand{\bs}{\boldsymbol}
\newcommand{\ms}{\mathsf}

\newcommand{\bpm}{\begin{pmatrix}}
\newcommand{\epm}{\end{pmatrix}}

\newcommand{\AEI}{\affiliation{Max Planck Institute for Gravitational Physics (Albert Einstein Institute), Am M\"uhlenberg 1, Potsdam 14476, Germany}}
\newcommand{\Perimeter}{\affiliation{Perimeter Institute for Theoretical Physics, Waterloo, Ontario N2L 2Y5, Canada}}
\newcommand{\ETH}{\affiliation{Institut für Theoretische Physik, Eidgenössische Technische Hochschule (ETH) Zürich, Switzerland}}

\begin{document}

\title{Test black holes, scattering amplitudes and perturbations of Kerr spacetime}

\author{Nils Siemonsen}\email{sinils@student.ethz.ch}\ETH\Perimeter

\author{Justin Vines}\email{justin.vines@aei.mpg.de}\AEI

\date{\today}

\begin{abstract}

It has been suggested that amplitudes for quantum higher-spin  massive particles exchanging gravitons lead, via a classical limit, to results for scattering of spinning black holes in general relativity, when the massive particles are in a certain way minimally coupled to gravity.  Such limits of such amplitudes suggest, at least at lower orders in spin, up to second order in the gravitational constant $G$, that the classical aligned-spin scattering function for an arbitrary-mass-ratio two-spinning-black-hole system can be obtained by a simple kinematical mapping from that for a spinning test black hole scattering off a stationary background Kerr black hole.  Here we test these suggestions, at orders beyond the reach of the post-Newtonian and post-Minkowskian results used in their initial partial verifications, by confronting them with results from general-relativistic ``self-force'' calculations of the linear perturbations of a Kerr spacetime sourced by a small orbiting body, here considering only results for circular orbits in the equatorial plane.  We translate between scattering and circular-orbit results by assuming the existence of a local-in-time canonical Hamiltonian governing the conservative dynamics of generic (bound and unbound) aligned-spin orbits, while employing the associated first law of spinning binary mechanics.  We confirm, through linear order in the mass ratio, some previous conjectures which would begin to fill in the spin-dependent parts of the conservative dynamics for arbitrary-mass-ratio aligned-spin binary black holes at the fourth-and-a-half and fifth post-Newtonian orders.

\end{abstract}


\maketitle


\section{Introduction}

The early successes of gravitational-wave (GW) astronomy have relied on highly accurate solutions to the binary black hole problem in general relativity (GR), obtained both from intensive numerical simulations (see e.g.\ \cite{Boyle:2019kee}) and, with overlapping domains of validity, from high-order calculations in the weak-field--slow-motion post-Newtonian (PN) approximation \cite{Blanchet:2013haa,Poisson:2014}, along with approaches to combining, interpolating and extrapolating these two crucial sources of information. The latter include effective-one-body models \cite{Buonanno99,Buonanno00,DJS,Barausse:2009xi,Damour:2014sva,Balmelli:2015zsa,Bini:2017xzy,Damour:2017zjx,Bini:2018ywr}, which have recently been linked to effective-one-body equivalences (simple maps from test-body motion in a stationary black hole background to arbitrary-mass-ratio two-body motion) which include and exactly resum infinite series of certain terms (with gauge-invariant information content) in the PN expansion, using \emph{only} the Schwarzschild or Kerr metric and mappings or identifications motivated by special-relativistic geometry and kinematics for asymptotic scattering states \cite{Vines:2016qwa,Damour:2016gwp,Vines:2017hyw,Vines:2018gqi}.

A pivotal challenge for future lower-frequency space-based GW observations will be the prediction of waveforms from extreme-mass-ratio binaries, with a few-solar-mass object spiraling into a supermassive black hole.  These lie in a regime currently inaccessible to both numerical relativity (due to the mass ratio) and PN treatments (due to the strong field) and require the use of black-hole perturbation theory within the ``self-force'' paradigm \cite{Gralla:2008fg,Poisson:2011nh,Harte:2014wya,Barack:2018yvs}, expanding about the limit of test-body motion in a stationary background spacetime --- most relevantly, the Kerr spacetime of a spinning black hole \cite{PhysRevLett.11.237,KerrSchild,Visser:2007fj,Teukolsky:2014vca}.  

A third perturbative approach to relativistic two-body dynamics, arguably situated somewhere in between the PN and self-force schemes, is the post-Minkowskian (PM) approximation, which assumes weak fields but places no restrictions on speeds (or on mass ratios), thus being most naturally applied to unbound scattering orbits, while still relevant for bound systems. The PN approximation can be obtained from a re-expansion in small velocities of the PM approximation. A recent resurgence of interest in fully PM computations and results has been associated with the prospect of applying advanced techniques for calculating quantum scattering amplitudes to the classical two-body problem; see e.g.\ references in \cite{Vines:2018gqi,Antonelli:2019ytb}.  While traditional classical methods had previously reached the 2PM level [$\mc O(G^2)$, where $G$ is Newton's gravitational constant], notably including Westpfahl's computation of the 2PM scattering-angle function for two monopolar point masses \cite{Westpfahl:1985}, a recent landmark has been the first calculation yielding analogous results at 3PM/two-loop order by Bern {et al.}\ \cite{Bern:2019nnu,Bern:2019crd},
using amplitudes computed with on-shell unitarity methods.  Whereas these results concern gravitational scattering of nonspinning bodies, our primary interest in this paper is in the dynamics of spinning bodies --- particularly, spinning black holes (BHs).


\subsection*{Spinning black holes from minimally coupled amplitudes}

It was suggested by Vaidya \cite{Vaidya:2014kza} that one can obtain classical Hamiltonians for the leading-order PN conservative dynamics of binary BHs in GR (through zeroth, first, second and fourth orders in the BHs' spins) from tree amplitudes for massive (spin-0, spin-1/2, spin-1 and spin-2) particles exchanging a graviton (respectively).  It had earlier been shown by Holstein and Ross \cite{Holstein:2008sx}, also at next-to-leading/one-loop order, that the classical linear-in-spin (or ``spin-orbit'') couplings are universal in the sense that the same results are obtained from both spin-1/2 and spin-1 massive particles (with an extension to spin-2 at leading order in \cite{Vaidya:2014kza}; see also work by Bjerrum-Bohr {et al.}\ e.g.\ in \cite{Bjerrum-Bohr:2013bxa}, by Damgaard et al.\ \cite{Damgaard:2019lfh}, and work by R\"udiger \cite{Rudiger:1981} on obtaining the [pole-dipole] Mathisson-Papapetrou-Dixon (MPD) equations \cite{Mathisson:2010,Papapetrou:1951pa,Tulczyjew:1959,Dixon:1979,Harte:2014wya,Dixon:2015vxa} from the Dirac equation in a curved spacetime). This is consistent with the classical fact that the dynamics at the pole-dipole level of the gravitational multipole expansion is universal for sufficiently isolated bodies in general relativity \cite{Mathisson:2010,Papapetrou:1951pa,Tulczyjew:1959,Dixon:1979,Gralla:2008fg,Harte:2014wya,Dixon:2015vxa}.  Vaidya \cite{Vaidya:2014kza} found furthermore that the leading-order spin-induced quadrupole couplings (which are quadratic in the spins) resulting from a minimally coupled massive spin-1 particle specifically match those appropriate for a BH.  From the spin-2 case, he obtained the BHs' spin-induced multipole series $\mc I_\ell+i\mc J_\ell=m(ia)^\ell$ \cite{Hansen:1974zz} up to the hexadecapoles, $\ell=4$, to quartic order in spin: $\mathcal{O}(a^4)$.  Here, $a=S/m$ is the ring radius of a BH with mass $m$ and spin (angular momentum) $S=ma$, with $c=1$.

Such classical-limit amplitude calculations were pushed to (significantly) higher spins at both tree and one-loop levels by Guevara \cite{Guevara:2017csg}.  Extending a novel treatment of the spin-0 case by Cachazo and Guevara \cite{Cachazo:2017jef} by employing the uniquely minimally coupled 3- and 4-point amplitudes for spinning massive particles meeting gravitons (or photons) proposed by Arkani-Hamed, Huang and Huang (AHH) \cite{Arkani-Hamed:2017jhn}, Guevara used on-shell methods to compute the fully special-relativistic classical limits of both (i) the tree-level amplitude for two minimally coupled massive spin-$s$ particles exchanging a graviton (or a photon) for \emph{arbitrary} spins $s$, and (ii) the corresponding one-loop amplitude for spins $s\le 2$ (for gravity, and for $s\le 1$ for electromagnetism) \cite{Guevara:2017csg}.  
It was shown by Guevara, Ochirov and Vines (GOV) \cite{Guevara:2018wpp}  how those amplitudes can be translated into a classical aligned-spin scattering-angle function, to all orders in spin [$\mc O(a^\infty)$] at $O(G^1)$, and through $\mc O(a^4)$ at $\mc O(G^2)$, in such a way that the result matches that for a binary BH in GR according to all known PN results \{through $\mc O(a^\infty)$ at the leading PN orders \cite{Vines:2016qwa,Siemonsen:2017yux} and through subsubleading PN orders through $\mc O(a^2)$ \cite{Blanchet:2013haa,Levi:2015ixa,Levi:2015msa,Porto:2016pyg,Levi:2016ofk,Levi:2018nxp}\} and all known PM results \{through $\mc O(a^\infty)$ at $\mc O(G^1)$ according to results of \cite{Vines:2017hyw}, and through $\mc O(a^1)$ at $\mc O(G^2)$ according to results of Bini and Damour \cite{Bini:2017xzy,Damour:2017zjx,Bini:2018ywr}\}, all for arbitrary mass ratios, as collected and analyzed in \cite{Vines:2018gqi}.  Beyond those orders, GOV \cite{Guevara:2018wpp} provided conjectural results up to $\mc O(G^2a^4)$ which have not yet been proven to correspond to binary BHs in general relativity.  The spinning amplitudes of \cite{Arkani-Hamed:2017jhn,Guevara:2017csg,Guevara:2018wpp} were also reproduced by Chung, Huang, Kim and Lee \cite{Chung:2018kqs}, who further translated them into contributions to the classical potential at subleading post-Newtonian orders, for the generic-spin case (not just for the aligned-spin case), while also proposing a parametrization of the 4-point ``Compton amplitude'' (two gravitons, two massive spin-$s$) for spins $s>2$ which would lead to $\mc O(G^2a^{>4})$ results; see also \cite{Chung:2019duq}.  

Subsequently, using and extending a formalism for extracting classical observables from on-shell amplitudes developed by Kosower, Maybee and O'Connell \cite{Kosower:2018adc}, the generalization to spinning particles was considered by Maybee et al.\ \cite{Maybee:2019jus}.  It was shown in \cite{Maybee:2019jus} through $\mc O(a^2)$, and then by Ochirov {et al.}\ in \cite{Guevara:2019fsj} through $\mc O(a^\infty)$ for both spins, that the arbitrary-spin tree amplitudes from \cite{Guevara:2017csg,Arkani-Hamed:2017jhn,Guevara:2018wpp} lead (by a well-motivated procedure) to the covariant scattering holonomy results for generic-spin binary BHs at $\mc O(G^1a^\infty)$ derived by one of the authors in \cite{Vines:2017hyw}, matching at $\mc O(G^1a^1)$ results of Bini and Damour \cite{Bini:2017xzy}.  A scalar-probe limit of the same generic-spin scattering holonomy, namely the $\mc O(G^1a^\infty)$ impulse (change in momentum) for a generic (weakly deflected) unbound geodesic in the Kerr spacetime, has also been produced from an elegant double copy of the amplitude for the analogous electromagnetic case, a test charge in the ``$\sqrt{\tr{Kerr}}$\,'' electromagnetic field in flat spacetime, by Arkani-Hamed, Huang and O'Connell \cite{Arkani-Hamed:2019ymq}; see also \cite{Moynihan:2019bor}.  Further perspectives on higher-spin-multipole effects along with double-copy constructions have been given by Johansson and Ochirov \cite{Johansson:2019dnu}, as well as by Bautista and Guevara \cite{Bautista:2019tdr,Bautista:2019evw} who also discussed the structure of radiative amplitudes with spin effects.  Double copies of spinning radiative processes have also been investigated by Goldberger, Li and Prabhu \cite{Goldberger:2017ogt,Li:2018qap}.  

\subsection*{Overview}

In this paper, we scrutinize the apparent duality between massive higher-spin quantum particles (in a classical limit) and classical spinning BHs, working through quadratic order in $G$ (to second order in the gravitational perturbations away from flat spacetime). We focus on relating aligned-spin-binary scattering-angle functions --- in particular the conjectural $\mc O(G^2a^4)$ binary-BH scattering function from GOV \cite{Guevara:2018wpp} constructed from the tree and one-loop amplitudes of Guevara \cite{Guevara:2017csg} using the 3- and 4-point amplitudes of AHH \cite{Arkani-Hamed:2017jhn} --- to information obtained in the context of gravitational self-force calculations --- in particular the high-order (8.5 and 8) PN expansions of the redshift \cite{Kavanagh:2016idg} and precession-frequency \cite{Bini:2018ylh} functions for circular equatorial orbits in a perturbed Kerr spacetime, obtained through linear order in the mass ratio by Kavanagh, Ottewill, and Wardell \cite{Kavanagh:2016idg} and Bini, Damour, Geralico, Kavanagh, and van de Meent \cite{Bini:2018ylh}.

To that end, we first review in Sec.~\ref{sec:chiX} results and conjectures for the aligned-spin binary-BH scattering function through $\mc O(G^2)$, particularly the conjecture \cite{Vines:2018gqi} (suggested by quantum amplitudes and classical-limit constructions \cite{Arkani-Hamed:2017jhn,Guevara:2017csg,Guevara:2018wpp}) that it is determined, via a simple mapping, by its spinning test-BH limit --- the limit where the mass ratio goes to zero while keeping the smaller BH's mass-rescaled multipole moments finite, in which it becomes a naked Kerr ring singularity of finite ring radius but negligible gravitational mass, a ``test BH.''

In Sec.~\ref{sec:TBH}, we parameterize the conservative dynamics of a test-BH in an arbitrary curved background in terms of dimensionless, constant ``Wilson coefficients'' appearing in its effective MPD equations of motion assumed to arise from an effective Hamiltonian action principle with the minimal effective degrees of freedom (translational and rotational), respecting appropriate general principles and symmetries.  All such coefficients contributing at linear order in curvature (or in $G$) are fixed by the considerations e.g.\ of \cite{Levi:2015msa,Vines:2016qwa,Siemonsen:2017yux,Vines:2017hyw} and matching to the stationary Kerr solution (at linear order), and we consider here the yet undetermined coefficients of conceivably relevant couplings which are quadratic in curvature and quartic in the test BH's spin $\sigma=S_\mr{test}/m_\mr{test}$.  We present results for a scattering function (an antiderivative of the scattering-angle function) for aligned-spin orbits of a test BH in the equatorial plane of a background Kerr spacetime, parametrized in terms of three dimensionless effective Wilson coefficients $\{C_\mr{4a},C_\mr{4c},C_\mr{4e}\}$ at $\mc O(G^2\sigma^4)$.  Further details of the calculations leading to these results (which can be accomplished e.g.\ by brute-force methods in Boyer-Lindquist coordinates, using the definitions and equations of motion discussed below) will be given in future work.

We discuss in Sec.~\ref{sec:Ham} the form of a reduced canonical Hamiltonian governing the local-in-time conservative dynamics of an aligned-spin binary BH system, for generic (bound and unbound) orbits, with most of the Hamiltonian's gauge freedom fixed by restricting attention to a certain (quasi-isotropic) class of phase space coordinates.  In Sec.~\ref{sec:circ} we discuss how the coefficients in the PM-PN and spin expansions of a quasi-isotropic aligned-spin Hamiltonian are related to the gauge-invariant coefficients in the expansions of the aligned-spin scattering function, and in turn to the coefficients in gauge-invariant functions characterizing circular aligned-spin orbits, namely the energy and angular momentum as functions of the orbital frequency.

In Sec.~\ref{sec:firstlaw}, we recall consequences of the first law of spinning binary mechanics, as developed or discussed by Le Tiec and collaborators e.g.\ in \cite{LeTiec:2011ab,Blanchet:2012at,Tiec:2014lba,Fujita:2016igj,LeTiec:2017ebm}, as applied to aligned-spin circular orbits.  This form of the first law relates changes in the gauge-invariant (center-of-mass-frame) binary parameters, the energy $E$, angular momenta (total $J=L+S_1+S_2$, [canoncial] orbital $L$, and spins $S_1$, $S_2$) and masses $m_1$, $m_2$, to the orbital frequency $\Omega$, the redshift invariants $z_1$, $z_2$, and the spin precession frequencies $\Omega_1$, $\Omega_2$.  
We present expressions for the redshift $z_1$ and precession frequency $\Omega_1$ for the smaller body $m_1$, to linear order in the mass ratio $q=m_1/m_2$, and to zeroth order in the small spin $S_1$, but to all orders in the larger body's dimensionless spin $\hat a=a_2/m_2=S_2/(Gm_2^2)$, through the leading and next-to-leading PN orders at each order in $\hat a$, through 5PN order --- specifically the expressions that result from matching via a Hamiltonian to the conjectural $O(G^2a^4)$ scattering function from the minimally coupled amplitudes.

We review in Sec.~\ref{sec:1SF} results for the same redshift and precession-frequency functions (of the orbital frequency, masses, and spins) through linear order in the mass ratio, obtained from first-order self-force calculations of the linear perturbations of a Kerr spacetime sourced by a small body in a circular orbit in the equatorial plane \cite{Kavanagh:2016idg,Bini:2018ylh}.

We compare the findings from Secs.~\ref{sec:firstlaw} and \ref{sec:1SF}, draw conclusions and discuss avenues for further study in Sec.~\ref{sec:conclude}.


\section{Spinning black hole scattering}

Considering a two-spinning-BH system in GR,
we restrict our attention to the conservative (time-symmetric) part of the classical dynamics, neglecting radiative or absorptive processes; this encompasses the complete orbital dynamics through $\mc O(G^2)$, at least at lower orders in spin. We furthermore consider only the aligned-spin configuration, in which both BHs' spins are constant and (anti)aligned with the orbital and total angular momenta.  
We review results and conjectures concerning aligned-spin scattering of two spinning BHs in Sec.~\ref{sec:chiX}, and we discuss an effective description of the motion of a spinning test BH in a general curved background, and then in the equatorial plane of a Kerr spacetime (with the test spin aligned with the axial symmetry direction), in Sec.~\ref{sec:TBH}, working through fourth order in the test spin $\sigma$.

\subsection{Elastic aligned-spin scattering functions for binary black holes: results and conjectures}\label{sec:chiX}

  It has been argued \cite{Vines:2017hyw,Vines:2018gqi}, following analogous arguments for the spinless case \cite{Damour:2016gwp,Damour:2017zjx}, that the complete gauge-invariant information content of a (perturbative) Hamiltonian (or Lagrangian, or set of equations of motion) governing the local-in-time conservative dynamics of an aligned-spin binary BH is contained in the aligned-spin scattering-angle function. This function is generally of the form
\be
\chi\big((m_1,a_1),(m_2,a_2),v,b\big),
\ee
giving the angle $\chi$ in the system's center-of-mass frame by which both BHs are deflected in a (weak) elastic scattering process, as a function of their constant masses $m_1$ and $m_2$ and (signed) ring radii $a_1=S_1/m_1$ and $a_2=S_2/m_2$, the relative velocity $v$ at infinity, and the ``proper'' or ``covariant'' impact parameter $b$ orthogonally separating the two BHs' asymptotic (both incoming or both outgoing) centroid worldlines defined by each BH's Fokker-Tulczyjew-Dixon supplementary condition \cite{Vines:2017hyw,Vines:2018gqi,Guevara:2018wpp}.  It was argued in \cite{Vines:2017hyw}, using the linearized Einstein equation and linear-level matching to the Kerr solution, that this function is well-defined through all orders in both BHs' spins at linear order in the coupling (and through all orders in $v$), and that it is given by
\be
\chi=\frac{GE}{v^2}\bigg(\frac{(1+v)^2}{b+a_1+a_2}+\frac{(1-v)^2}{b-a_1-a_2}\bigg)+\mc O(G^2),
\ee
where $E$ is the total energy,
\be
E^2=(p_1+p_2)^2=m_1^2+m_2^2+2m_1m_2\gamma,
\ee
with the relative Lorentz factor
\be
\gamma=\frac{1}{\sqrt{1-v^2}}=\frac{p_1\cdot p_2}{m_1m_2}
\ee
at infinity, in terms of the asymptotic incoming (or outgoing) 4-momenta $p_1$ and $p_2$.  

To more efficiently express results and conjectures for the aligned-spin binary-BH scattering-angle function $\chi$ at $\mc O(G^2)$, let us use an aligned-spin ``scattering function'' $X((m_1,a_1),(m_2,a_2),v,b)$ such that
\be\label{defX}
\chi=\frac{\doe X}{\doe b},
\ee
expanded, with $X^{(k)}= \mathcal{O}(G^k)$, as
\be
X=X^\mr{(1)}+X^\mr{(2)}+\mathcal{O}(G^3),
\ee
given through linear order by
\begin{alignat}{3}\label{X1}
X^\mr{(1)}=\frac{GE}{v^2}\sum_\pm (1\pm v)^2\log\frac{b\pm a_1\pm a_2}{\mr{const.}}.
\end{alignat}

According to the $\mc O(G^2)$ spinless results of Westpfahl \cite{Westpfahl:1985} and the $\mc O(G^2)$ spin-orbit results of Bini and Damour \cite{Bini:2017xzy,Bini:2018ywr}, $X^\mr{(2)}$ through $\mc O(a^1)$ is universal (applying not only to BHs, but to any sufficiently isolated bodies) and is given by
\begin{alignat}{3}\label{linearinspinresult}
X^\mr{(2)}&=\frac{\pi G^2E}{b}\bigg[{-}\frac{3}{4}(m_1+m_2)\frac{4+v^2}{v^2}
\nnm\\\nnm
&\quad+\frac{1}{2b}\Big(m_1(4a_1+3a_2)+m_2(4a_2+3a_1)\Big)\frac{2+3v^2}{v^3}\bigg]
\\
&\quad+\mc O(a^2).\phantom{\bigg]}
\end{alignat}
Note that this result also uniquely determines the generic-spin (not just aligned-spin) results at $\mc O(G^2a^1)$ \cite{Bini:2017xzy,Bini:2018ywr}.

Let us consider, from these results, or from results at higher orders in spin for BHs, taking a spinning-test-body limit (at fixed $v$ and $b$). This entails taking the mass of the smaller body to zero, \mbox{$m_1\to 0$}, while preserving $p_1^2/m_1^2=1$ at infinity, and while keeping fixed the test body's mass-rescaled spin, $a_1\to \sigma$, the test spin. We thereby obtain an aligned-spin scattering function $X_\tr{\underline{t}}$ for a spinning \underline{t}est body of ``ring radius'' $\sigma$ in the equatorial plane of a background Kerr spacetime of mass ($E\to$) $m_2\to m$ and ring radius $a_2\to a$,
\be
X_\mr t(m,a,\sigma,v,b)=X\big((0,\sigma),(m,a),v,b\big).
\ee
This limit is well-defined (and nontrivial, since the outcome still depends on $\sigma$) for the arbitrary-mass-ratio results (\ref{X1}) and  (\ref{linearinspinresult}) through $\mc O(G^2a^1)$.  Furthermore, as noted in \cite{Vines:2018gqi}, the arbitrary-mass-ratio result $X$ is recovered from its spinning-test-body limit $X_\mr t$ via the mapping
\begin{alignat}{3}\label{EOBmap}
&X\big((m_1,a_1),(m_2,a_2),v,b)\phantom{\bigg]}
\nnm\\\nnm
&=\frac{E}{m_1+m_2}\bigg[\frac{m_1}{m_1+m_2}X_\mr t(m_1+m_2,a_1,a_2,v,b)
\\\nnm
&\qquad\qquad\quad+\frac{m_2}{m_1+m_2}X_\mr t(m_1+m_2,a_2,a_1,v,b)\bigg]
\\
&\quad+\mc O(G^2a^{\ell+1})+\mc O(G^3)\phantom{\bigg]}
\end{alignat}
with $\ell=1$ according to the result (\ref{linearinspinresult}) from Bini and Damour \cite{Bini:2018ywr}.

\begin{figure*}[t]
\includegraphics[width=1\textwidth]{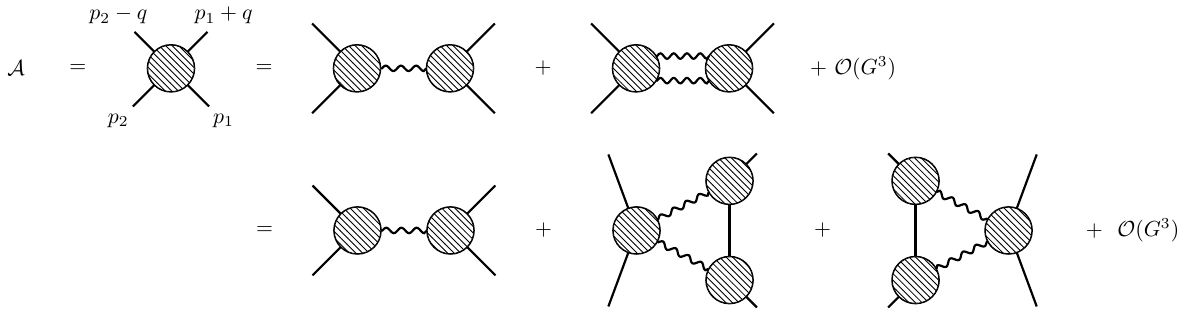}
\caption{ The tree- and 1-loop-level structure of the massive spin-s particle-graviton on-shell scattering amplitude $\mathcal{A}$, where $q$ is the momentum transfer. When moving from the amplitude $\mathcal{A}$ to a classical scattering-angle function $\chi$, the momentum transfer $q$ of $\mathcal{A}$ translates into the impact parameter $b$ of $\chi$. For higher-spin particles the amplitude structure $\mathcal{A}=\mathcal{A}^\text{tree}+\mathcal{A}^\text{1-loop}+\mathcal{O}(G^3)$ suggests the mapping \eqref{EOBmap}. To see this, we follow the arguments in \citep{Vines:2018gqi}, and point out that the structure $\mathcal{A}^\text{1-loop}=\mathcal{A}^\text{1-loop}_\triangleleft+\mathcal{A}^\text{1-loop}_\triangleright$ yields the following decomposition of the scattering-angle function: $\chi[\mathcal{A}]=\chi[\mathcal{A}^\text{tree}]+\chi[\mathcal{A}^\text{1-loop}_\triangleleft]+\chi[\mathcal{A}^\text{1-loop}_\triangleright]+\mathcal{O}(G^3)$, where $\chi[\mathcal{A}^\text{1-loop}_\triangleleft]=E m_2 f_\triangleleft$ and $\chi[\mathcal{A}^\text{1-loop}_\triangleright]=E m_1 f_\triangleright$. Crucially, $f_\triangleright\equiv f_\triangleleft$, and $f_\triangleright$ is independent of both $m_1$ and $m_2$. Therefore, even in the extreme-mass-ratio limit (i.e., $m_1\rightarrow 0$), the \textit{full arbitrary mass-ratio} information can be recovered. The validity of this decomposition is discussed in the text, and in \cite{Vines:2018gqi}. 
}
\label{triangleamp}
\end{figure*}

It was shown in \cite{Vines:2018gqi} that the extrapolation of the $\mc O(G^2)$ binary-black-hole ``effective-one-body aligned-spin scattering-angle mapping'' (\ref{EOBmap}) to $\ell=2$, to quadratic order in the spins, is fully consistent with the arbitrary-mass-ratio subsubleading-PN $\mc O(a^2)$ results of Levi and Steinhoff \cite{Levi:2015ixa,Levi:2015msa,Levi:2016ofk,Levi:2018nxp} (and fully determines the aligned-spin results through subleading orders), together with the results of Bini et al.\ \cite{Bini:2017pee} for the unique (minimal) MPD dynamics for an effective quadrupolar test-BH \cite{Porto:2008jj,Steinhoff:2010zz,Blanchet:2013haa,Marsat:2014xea,Steinhoff:2014,Porto:2016pyg,Levi:2018nxp} undergoing aligned-spin scattering in a background Kerr spacetime. 
The results of \cite{Bini:2017pee} lead to an $\mc O(G^2)$ test-BH-in-Kerr scattering function $X_t^{(2)}$, to quadratic order in the test spin $\sigma$, but to all orders in the background Kerr spin $a$ (to all orders in $v$), given, modulo the apparent freedom in its definition \eqref{defX}, by
\begin{alignat}{3}\label{quadrupolarTBH}
X^{(2)}_\mr t=\frac{\pi}{2}\frac{(Gm)^2}{v^4a^2}\bigg[v^4b&-\frac{(vb-a)^4}{(b^2-a^2)^{3/2}}
\nnm\\\nnm
&-\sigma \frac{\doe}{\doe b}\bigg(\frac{(vb-a)^3(b-va)}{(b^2-a^2)^{3/2}}\bigg)
\\\nnm
&-\frac{\sigma^2}{2}\frac{\doe^2}{\doe b^2}\bigg(\frac{(vb-a)^2(b-va)^2}{(b^2-a^2)^{3/2}}\bigg)
\bigg]
\\
\phantom{\bigg[}+\mc O(\sigma^3).\quad\;\;&
\end{alignat}
The resulting scattering angle $\chi_\mr t=\doe X_\mr t/\doe b$ is finite as $a\to0$ in spite of the appearance of $a^2$ in the overall denominator.  The expansion of this result to quadratic order in either spin, $\mc O(a,\sigma)^2$, matches Eq.~(5.5) of \cite{Vines:2018gqi}.

The mapping (\ref{EOBmap}) with $\ell=1$, applied to the $\mc O(a,\sigma)^1$ truncation of the result (\ref{quadrupolarTBH}) for a spinning test BH in Kerr, correctly yields the $\mc O(G^2a^1v^\infty)$ arbitrary-mass spin-orbit results (\ref{linearinspinresult}) from Bini and Damour \cite{Bini:2018ywr}. The extrapolation of the mapping (\ref{EOBmap}) to $\ell=2$, to quadratic order in spins (to all orders in $v$), applied to the  $\mc O(a,\sigma)^2$  truncation of the quadrupolar test-BH-in-Kerr result (\ref{quadrupolarTBH}), yields the first, simplest and most verified of the conjectural results for the arbitrary-mass aligned-spin binary-black-hole scattering function at $\mc O(G^2)$, suggested in \cite{Vines:2018gqi} or \cite{Guevara:2018wpp}, which warrant scrutiny --- and it is the one which our considerations in this paper will not be able to probe beyond the levels where it has already been shown to concur with the arbitrary-mass subsubleading-PN $\mc O(a^2)$ results \cite{Levi:2015ixa,Vines:2018gqi}; we will instead be able to probe higher PN orders at higher orders in spin, $\mc O(a^{\ell>2})$.

The classical-limit higher-spin amplitude constructions of GOV \cite{Guevara:2018wpp} --- based on the building-block amplitudes presented by Arkani-Hamed, Huang and Huang \cite{Arkani-Hamed:2017jhn}, which were first stitched by Guevara \cite{Guevara:2017csg} into one- and two-graviton exchanges between two massive spinning particles --- suggest that the mapping (\ref{EOBmap}) will continue to hold for $\ell=4$, through quartic order in spins.  See Fig.~\ref{triangleamp}. It is furthermore suggested \cite{Guevara:2018wpp} that the associated $\mc O(G^2a^\infty\sigma^4)$ hexadecapolar-test-BH-in-Kerr scattering function is given by the most naive conceivable extrapolation to $\mc O(\sigma^4)$ of the pattern emerging in the expression (\ref{quadrupolarTBH}):
\bse\label{GOVTBH}
\begin{alignat}{3}\label{GOVTBHa}
X^{(2)}_\mr{t_\mr{GOV}}&=\frac{\pi}{2}\frac{(Gm)^2}{a^2}\bigg(b-\sum_{\ell=0}^4\frac{\sigma^\ell}{\ell!}\frac{\doe^\ell}{\doe b^\ell}\frac{(vb-a)^{4-\ell}(b-va)^\ell}{v^4(b^2-a^2)^{3/2}}\bigg)
\nnm\\
&\quad+\mc O(\sigma^5)\phantom{\Big]}
\\
\py\nnm
\\
&=\frac{\pi}{2v^4}\oint_\Gamma\frac{\mr dz}{2\pi i}\frac{(1-vz)^4}{(z^2-1)^{3/2}}\bigg(b-z a-\frac{z-v}{1-vz}\sigma\bigg)^{-1}
\nnm\\\label{contour}
&\quad+\mc O(\sigma^5).\phantom{\Big]}
\end{alignat}
\ese
An appropriate contour $\Gamma$ in (\ref{contour}) is discussed in \cite{Guevara:2018wpp}. The extrapolation of (\ref{GOVTBHa}) to $\ell=5$ is likely invalid for our purposes, as the naive extrapolations to a spin-5/2 particle of the amplitudes on which it is based are known to be pathological (or at least nonlocal) \cite{Arkani-Hamed:2017jhn,Guevara:2018wpp,Chung:2018kqs,Chung:2019duq}.

\subsection{Minimal effective test-black-hole dynamics}\label{sec:TBH}

We report here that the test-BH limit (\ref{GOVTBH}) of the conjectural $\mc O(G^2\sigma^4)$ result from GOV \cite{Guevara:2018wpp} [which is (\ref{GOVTBH}) plugged into (\ref{EOBmap}) with $\ell=4$] coincides through $\mc O(\sigma^3)$ with the scattering function obtained from the (seemingly) unique cubic-in-test-spin minimal MPD effective-test-BH dynamics (consistent with the appropriate symmetries and limits and the Kerr solution, as discussed e.g.\ by Marsat \cite{Marsat:2014xea}) as applied to aligned-spin scattering in a Kerr background through $\mc O(Gm)^2$.  Details of the calculation leading to this and the following results will be given in future work.

The minimal MPD dynamics to which we are referring is minimal in the sense that the higher-order relativistic multipole moments (as defined by Dixon \cite{Dixon:1979,Harte:2014wya,Dixon:2015vxa}) depend only on the momentum $p_\mu$ (the dynamical gravitational monopole) and the spin angular momentum 2-form $S_{\mu\nu}$ (the dynamical gravitational dipole), and on the location in spacetime (covariantly), with any other conceivable degrees of freedom having been integrated out (to the extent that this can be well-defined).  The dynamics can be encoded in an effective action principle \cite{Vines:2017hyw,Vines:2016unv,Porto:2008jj,Steinhoff:2010zz,Blanchet:2013haa,Marsat:2014xea,Steinhoff:2014,Porto:2016pyg,Levi:2018nxp},\footnote{
One form of an appropriate action functional, one which directly leads to the form (\ref{minimalMPD}) of the minimal MPD equations of motion,
$$
\int d\tau\bigg[p_\mu\dot z^\mu +\frac{1}{2}S_{\mu\nu}\Omega^{\mu\nu}+\beta^\mu S_{\mu\nu}p^\nu+\frac{\alpha}{2}\Big(p^2-\mc M^2(z,p,S)\Big)\bigg],
$$
is discussed e.g.\ in Appendix B of \cite{Vines:2017hyw} and in \cite{Vines:2016unv}.  (See also \cite{Porto:2008jj,Steinhoff:2010zz,Blanchet:2013haa,Marsat:2014xea,Steinhoff:2014,Porto:2016pyg,Levi:2018nxp}.)  The independent dynamical variables to be varied here are the worldline $x=z(\tau)$, the momentum $p_\mu(\tau)$ and the spin (angular momentum) 2-form $S_{\mu\nu}(\tau)$ along the worldline, and auxiliary fields along the worldline, the body-fixed orthonormal tetrad $\Lambda_a{}^\mu(\tau)$ [with global internal Lorentz symmetry $\Lambda_a{}^\mu(\tau)\leftrightarrow L_a{}^b\Lambda_b{}^\mu(\tau)$] determining the angular velocity tensor $\Omega^{\mu\nu}=\Lambda_a{}^\mu\dfrac{\mr D\Lambda^{a\nu}}{\mr d\tau}$, and the Lagrange multipliers $\beta^\mu$ and $\alpha$ enforcing respectively the Tulczyjew condition (\ref{Tulcz}) and the dynamical mass-shell constraint (\ref{massshell}).  In (\ref{minimalMPD}), $\dfrac{\mr D}{\mr Dz^\mu}=\nabla^\mr{h}_\mu$ is the horizontal covariant derivative with respect to the worldline point $z$, parallelly
transporting $p$ and $S$ between neighboring points.
} 
leading to equations of motion equivalent to the MPD equations, transport equations for $p_\mu$ and $S_{\mu\nu}$ along a centroid worldline $x=z(\tau)$ with tangent $\dot z^\mu=\mr dz^\mu/\mr d\tau$,
\begin{alignat}{3}\label{minimalMPD}
&\frac{\mr D}{\mr d\tau}p_\mu+\frac{1}{2}R_{\mu\nu\kappa\lambda}\dot z^\nu S^{\kappa\lambda}=\frac{p\cdot\dot z}{2}\frac{\mr D}{\mr D z^\mu}\log\mc M^2,
\\\nnm
&\frac{\mr D}{\mr d\tau}S^{\mu\nu}-2p^{[\mu}\dot z^{\nu]}=p\cdot\dot z\bigg(p^{[\mu}\frac{\doe}{\doe p_{\nu]}}+2S^{[\mu}{}_\rho\frac{\doe }{\doe S_{\nu]\rho}}\bigg)\log \mc M^2,
\end{alignat}
along with the Tulczyjew condition,
\be\label{Tulcz}
p_\mu S^{\mu\nu}=0.
\ee  
The terms on the right-hand sides of (\ref{minimalMPD}), representing higher-multipole couplings, are given (under the assumption of the existence of a minimal effective action) in terms of the ``dynamical mass function'' $\mc M(p,S,z)$ determining the magnitude $\sqrt{p^2}$ of the momentum $p_\mu$ as a function of (i) its direction,
\be
u^\mu=\frac{p^\mu}{\sqrt{p^2}},
\ee
(ii) the spin $S_{\mu\nu}$, or equivalently the rescaled\footnote{The factor $m$ by which the spin is rescaled here is the ``bare'' rest mass, which is generally a function $m(S^2)$ of the conserved spin length $S^2=\frac{1}{2}S_{\mu\nu}S^{\mu\nu}=m^2(-\sigma^2)$, appearing in (\ref{GOVmcM}) as $\mc M^2=m^2+\mc O(R)$ where $\mc O(R)$ denotes terms with one or more powers of the curvature tensors, which are assumed to vanish at infinity.} spin vector,
\be
\sigma^\mu=\frac{1}{2m}\epsilon^{\mu\nu\kappa\lambda} u_\nu S_{\kappa\lambda}
\quad\Leftrightarrow\quad
S_{\mu\nu}=m\,\epsilon_{\mu\nu\kappa\lambda}u^\kappa \sigma^\lambda,
\ee
and (iii) the spacetime location $x=z$ (only through the covariant metric and its covariant curvature tensors, the Riemann tensor and its symmetrized covariant derivatives, evaluated locally along the worldline),
\begin{alignat}{3}\label{massshell}
p^2&=\mc M^2(p,S,z)
\\\nnm
&=\mc M^2\Big(u_\mu,\sigma_{\mu},g_{\mu\nu}(z),\{R_{\mu\nu\kappa\lambda;(\rho_1\ldots \rho_{\ell-2})}(z)\}\Big).
\end{alignat}
Taking this function to be given --- presuming to have all terms that will matter for a consistent spinning-test-BH limit through $\mc O(G^2\sigma^4)$ --- by\footnote{A consistent spinning test BH limit is assumed here to respect parity and time-reversal, and to be such that the squared dynamical mass $\mc M^2=p^2$ divided by the squared bare rest mass $m^2$ is independent of the bare rest mass, with any dimensioned scale for the test BH dynamics determined only by its ring radius $\sqrt{-\sigma_\mu\sigma^\mu}$.  The mass should ``scale out'' for a ``test BH.''
Note, ${}^{*\!}R_{\mu\nu\kappa\lambda}=\frac{1}{2}\epsilon_{\mu\nu}{}^{\rho\sigma}R_{\rho\sigma\kappa\lambda}$ is the (left or right) dual of the vacuum Riemann (Weyl) tensor, and ${}^{*\!}R_{\mu\nu\kappa\lambda;\rho}$ is its covariant derivative.}
\begin{alignat}{3}\label{GOVmcM}
&\mc M^2_{{}_\mr{GOV}}=m^2+2m^2 u^\mu u^\nu \sigma^{\rho_1}\sigma^{\rho_2} \bigg({-}\frac{1}{2!}R_{\mu\rho_1\nu\rho_2}
\\\nnm
&\;\;+\frac{1}{3!} {}^{*\!}R_{\mu\rho_1\nu\rho_2;\rho_3} \sigma^{\rho_3}+\frac{1}{4!} R_{\mu\rho_1\nu\rho_2;\rho_3\rho_4} \sigma^{\rho_3}\sigma^{\rho_4}\bigg)+\mc O(\sigma^5),
\end{alignat}
leads, in a Kerr background, to an aligned-spin scattering function  which precisely matches the $\mc O(G^2\sigma^4)$ (spinning-test-BH-limit-)GOV result (\ref{GOVTBH}) --- as well as precisely matching the $\mc O(G^1\sigma^\infty)$ result (\ref{X1}) matched to linearized Kerr \cite{Vines:2017hyw} up to $\mc O(\sigma^4)$, which already fixes all of the terms in (\ref{GOVmcM}) and their coefficients \cite{Levi:2015msa,Vines:2017hyw,Chung:2018kqs}.

However, (\ref{GOVmcM}) does not represent the unique set of consistent test-BH couplings contributing up to $\mc O(G^2\sigma^4)$, allowed by general principles in a minimal MPD effective-action approach (in an assumed vacuum background), and consistent with linear-level matching to the Kerr solution.  Rather, there arise at fourth order in the test spin $\sigma$ a set of conceivable relevant couplings with coefficients $C_{4..}$ not fixed by the aforementioned considerations.  Six of the seven couplings are quadratic in the curvature $R$ and contribute at the same orders as the $\mc O(G^2a^4)$ terms already seen above in (\ref{GOVTBH}) [with the latter arising from interactions among the up-to-linear-in-curvature terms in (\ref{GOVmcM}) and (\ref{minimalMPD})].  The possible additional $\mc O(R^2\sigma^4)$ couplings [and a seventh $\mc O(R\sigma^4)$ coupling] can be organized in terms of the symmetric-trace-free electric and magnetic parts, with respect to $u^\mu$, of the vacuum Riemann (Weyl) tensor (the ``tidal tensors''),
\be
\mc E_{\mu\nu}=R_{\mu\kappa\nu\lambda}u^\kappa u^\lambda,
\qquad
\mc B_{\mu\nu}={}^{*\!}R_{\mu\kappa\nu\lambda}u^\kappa u^\lambda,
\ee
as follows:
\begin{alignat}{3}\label{dmcM4}
\frac{1}{m^2}\,\delta(\mc M^2)_4&=\phantom{+}C_\mr{4A}\,(\mc E_{\mu\nu}\sigma^\mu \sigma^\nu)^2
\\\nnm
&\quad+C_\mr{4B}\,\mc E_{\mu\lambda}\mc E_{\nu}{}^\lambda \sigma^\mu \sigma^\nu (-\sigma^2)
\\\nnm
&\quad+C_\mr{4C}\,\mc E_{\kappa\lambda}\mc E^{\kappa\lambda} (-\sigma^2)^2\phantom{\frac{1}{2}}
\\\nnm
&\quad+C_\mr{4D}\,(\mc B_{\mu\nu}\sigma^\mu \sigma^\nu)^2
\\\nnm
&\quad+C_\mr{4E}\,\mc B_{\mu\lambda}\mc B_{\nu}{}^\lambda \sigma^\mu \sigma^\nu (-\sigma^2)\phantom{\frac{1}{2}}
\\\nnm
&\quad+C_\mr{4F}\,\mc B_{\kappa\lambda}\mc B^{\kappa\lambda} (-\sigma^2)^2,
\\\nnm
&\quad+C_\mr{4G}\,\ddot{ \mc E}_{\mu\nu}\sigma^\mu \sigma^\nu (-\sigma^2),\phantom{\frac{1}{2}}
\end{alignat}
with $-\sigma^2=-\sigma_\mu\sigma^\mu$ (not to be confused with the oriented radius $\sigma=\pm\sqrt{-\sigma^2}$ in the arguments of the aligned-spin scattering functions).  At least concerning their contributions for aligned-spin scattering in a Kerr background, these terms are degenerate; the resultant $\mc O(G^2\sigma^4)$ scattering function depends only on the three linear combinations
\begin{alignat}{3}\label{C4as}
C_\mr{4a}&=C_\mr{4A}+C_\mr{4B},
\qquad
C_\mr{4c}=C_\mr{4C},\phantom{\frac{1}{2}}
\\\nnm
C_\mr{4e}&=C_\mr{4E}+\frac{C_\mr{4F}}{2},
\end{alignat}
of the seven constant dimensionless ``Wilson coefficients'' $C_\tr{4..}$ in (\ref{dmcM4}).  The term multiplying $C_\mr{4D}$ is identically zero for aligned spins, and the linear-in-curvature $C_\mr{4G}$ term (with $\ddot{ \mc E}_{\mu\nu}:=R_{\rho_1\mu\rho_2\nu;\rho_3\rho_4}u^{\rho_1}u^{\rho_2}u^{\rho_3}u^{\rho_4}$) contributes zero to the scattering function at $\mc O(G^{1,2}\sigma^4)$, first yielding nonzero contributions at $\mc O(G^2\sigma^5)$.  The possible contributions to the 
$\mc O(G^2\sigma^4)$ test-BH-in-Kerr aligned-spin scattering function $X_\mr t^{(2)}$ in addition to those in (\ref{GOVTBH}), resulting from adding $m^2$ $\times$ (\ref{dmcM4}) to (\ref{GOVmcM}), are
\begin{alignat}{3}\label{deltaX4}
&\delta(X_\mr t^{(2)})_4=\sigma^4\frac{\doe^4}{\doe b^4}\Bigg(\frac{\pi (Gm)^2}{2^8v^2(1-v^2)a^4\sqrt{b^2-a^2}}\bigg\{
\\\nnm
&C_\mr{4a}\Big[(8-8v^2+3v^4)b^4-4vba[(4-v^2)b^2-(1-4v^2)a^2]
\\\nnm
&\quad-2(4-17v^2+4v^4)b^2a^2+(3-8v^2+8v^4)a^4\Big]
\\\nnm
&+2C_\mr{4c}\Big[(8-16v^2+11v^4)b^4-12vba(v^2b^2+a^2)
\\\nnm
&\quad-2(8-25v^2+8v^4)b^2a^2
+(11-16v^2+8v^4)a^4\Big]
\\\nnm
&-C_\mr{4e}\Big[(8-16v^2+5v^4)b^4+12vba(v^2b^2+a^2)
\\\nnm
&\quad-2(8-7v^2+8v^4)b^2a^2
+(5-16v^2+8v^4)a^4\Big]\bigg\}\Bigg).
\end{alignat}
Assuming that we have fully parametrized the freedom in the conceivable consistent $\mc O(G^2a^4)$ test-BH-in-Kerr aligned-spin scattering functions (arising from a minimal MPD effective action and consistent with linear-level matching to Kerr), it is notable that this result is still quite constrained, with only three free (constant, dimensionless) parameters, while consisting of quartic polynomials in $(b,a)$ with coefficients being up-to-quadratic polynomials in $v^2$.  

It is also notable that the term multiplying each of the effective Wilson coefficients $C_\mr{4a}$, $C_\mr{4c}$ and $C_\mr{4e}$, individually, diverges in the ultrarelativistic limit $v\to1$, due to the overall factor of $1-v^2$ in the denominator in (\ref{deltaX4}); the contributions to the function $X$ as well as to the angle $\chi=\doe X/\doe b$ diverge as $v\to1$ (while the $\chi$ contribution from each $C_\mr{4..}$ term is finite as $a\to0$).  Contrastingly, the $\mc O(G^1a^\infty)$ results (\ref{X1}), the $\mc O(G^2a^1)$ results (\ref{linearinspinresult}) and the conjectural $\mc O(G^2a^4)$ results (\ref{GOVTBH}) are all finite as $v\to1$.  From (\ref{deltaX4}), we have the scattering-angle contribution
\begin{alignat}{4}\label{chi4div}
\delta(\chi_\mr t^{(2)})_4&=-\frac{315\pi (Gm)^2(5b-4a)\sigma^4}{256(b+a)^4(b^2-a^2)^{3/2}}\frac{C_\mr{4a}+2C_\mr{4c}+C_\mr{4e}}{1-v^2}
\nnm\\
&\quad+\mc O(1-v^2)^0
\end{alignat}
at the diverging order in the ultrarelativistic limit.  If we were to demand that the $\mc O(\sigma^4)$ contribution to the scattering angle be finite as $v\to1$, we  would set
\begin{align}
C_\mr{4a}+2C_\mr{4c}+C_\mr{4e}\;\ddot=\;0
\label{WCeq1}
\end{align}
and be left with a 2-parameter family of test-BH-in-Kerr scattering functions at $\mc O(G^2\sigma^4)$.

If the thus-constructed test-BH-in-Kerr scattering function is to describe the spinning-test-body limit of the $\mc O(G^2a^4)$ aligned-spin binary-black-hole scattering function (assuming for now that such a thing is well defined), how can the remaining Wilson coefficients be determined?  Is there a general principle from which we should expect the scattering angle $\chi_\mr t(m,a,\sigma,v,b)$ to be finite as $v\to1$ at fixed $(m,a,\sigma,b)$?  As we have so far encountered constraints coming from matching to the Kerr solution only at the linearized-off-of-flat-spacetime level (as implemented e.g.\ in \cite{Vines:2017hyw}), are there further constraints coming from matching to the stationary Kerr solution at $\mc O(Gm)^2$?  While that is possible, it is also possible that a complete understanding of ``the $\mc O(Gm)^2$ level'' (if this is well defined) will require analysis of perturbations of the Kerr spacetime.

\py

\section{Hamiltonians for generic aligned-spin orbits and invariants for circular orbits}\label{sec:genericorbits}

We now turn to translating the scattering functions discussed above into dynamical information describing bound (as well as unbound) orbits of aligned-spin binary BHs. We first discuss a class of aligned-spin reduced canonical Hamiltonians in quasi-isotropic ``gauges'' (choices of phase-space coordinates), which are sufficient to describe generic (bound and unbound) orbits under the assumption of the existence of a local-in-time Hamiltonian or action principle for the aligned-spin binary-BH conservative dynamics. Secondly, we construct the total conserved energy $E$ and angular momentum $J=L+S_1+S_2$ as functions of the masses and spins and the orbital frequency $\Omega$ for circular orbits of aligned-spin binaries.

We consider the conjecture ($\dot=$) that the EOB scattering-angle mapping (\ref{EOBmap}) holds with $\ell=4$, so that the aligned-spin binary-black-hole scattering function $X((m_1,a_1),(m_2,a_2),v,b)$ is determined through $\mc O(G^2a^5)$ by its (assumed well-defined) spinning-test-BH limit $X_\mr t(m,a,\sigma,v,b)$.  We further conjecture that this $X_\mr t$ takes the form discussed in the previous section, unique up to $\mc O(G^2a^\infty\sigma^3)$, and parametrized by the effective Wilson coefficients $C_\mr{4..}$ through $\mc O(G^2a^\infty\sigma^4)$.

\subsection{Canonical local-in-time Hamiltonians for spinning binaries
 in unbound and bound orbits}\label{sec:Ham}

More generally, given some scattering-angle function $\chi(v,b;m_\ms a,a_\ms a)=\frac{\doe}{\doe b} X(v,b;m_\ms a,a_\ms a)$, with $\ms a=1,2$, as discussed in \cite{Vines:2018gqi}, we can compute from it an aligned-spin canonical Hamiltonian
\be
H(R,P_R,L;m_\ms a,S_\ms a),
\ee 
governing the conservative dynamics of both unbound and bound aligned-spin orbits according to
\begin{alignat}{3}\label{Hampolar}
\begin{aligned}
\dot R&=\dfrac{\doe H}{\doe P_R},
\qquad&
\dot{P}_R&=-\dfrac{\doe H}{\doe R},
\\
\dot\phi&=\dfrac{\doe H}{\doe L},
&
\dot{L}&=-\dfrac{\doe H}{\doe\phi}=0,
\end{aligned}
\end{alignat}
to whatever levels in a dual PM-PN expansion which knowledge of the scattering function allows, as follows.  
Suppressing the dependence on the masses $m_\ms a=\{m_1,m_2\}=\{m\}$, with sums over $\ms a=1,2$ implied here for the spins
\be
S_\ms a=\{S_1,S_2\}=\{m_1a_1,m_2a_2\},
\ee
and with
$
P^2=P_R^2+\dfrac{L^2}{R^2},
$
we pose the quasi-isotropic-gauge canonical Hamiltonian \cite{Vines:2018gqi} ansatz
\bse\label{Ham}
\begin{alignat}{3}\label{qiHam}
&H(R,P_R,L;\{m,S\})=H_0(P^2)
\\\nnm
&\quad+\sum_k^{1,2}\frac{G^k}{R^k}\Bigg[\sum_{\ell}^{0,2,4}H_{k\ell}^{\ms a_1\ldots \ms a_\ell}(P^2)\frac{S_{\ms a_1}...\, S_{\ms a_\ell}}{R^\ell}
\\\nnm
&\qquad+\sum_{\ell}^{1,3,5}H_{k\ell}^{\ms a_1\ldots \ms a_\ell}(P^2)\frac{LS_{\ms a_1}...\, S_{\ms a_\ell}}{R^{\ell+1}}+\mc O(S^7)\Bigg]+\mc \mc O(G^3),
\end{alignat}
with
\be
H_{k\ell}^{\ms a_1\ldots\ms a_\ell}(P^2)=\sum_{n=0}^\infty (P^2)^n H_{k\ell n}^{\ms a_1\ldots\ms a_\ell}(\{m\}),
\ee
\ese
where we include $S^5$ terms (for our purposes below) which can be partially determined from the conjectural $\mc O(G^2a^\infty\sigma^4)$ results (in particular the $S_1^4S_2$ and $S_1S_2^4$ terms but not the $S_1^5$ and $S_2^5$ terms).

The scattering angle $\chi$ is determined from the Hamiltonian by solving $E=H(R,P_R,L)$ for $P_R(R,E,L)$ and integrating \cite{Damour:2017zjx}
\be
\pi+\chi(E,L)=-2\int_{R_\mr{min}}^\infty dR\,\frac{\doe}{\doe L}P_R(R,E,L),
\ee
where the turning point $R_\mr{min}$ is the largest root of $P_R(R,E,L)=0$.  The change of variables from the physical center-of-mass-frame energy $E$ and canonical orbital angular momentum $L$ to the asymptotic relative velocity $v$ and ``proper'' impact parameter $b$ is accomplished with special-relativistic kinematics at infinity \cite{Vines:2017hyw,Vines:2018gqi} (see also \cite{Bini:2017xzy,Bini:2018ywr})
\begin{alignat}{3}
E^2&=m_1^2+m_2^2+\frac{2m_1m_2}{\sqrt{1-v^2}},
\\\nnm
L&=\frac{m_1m_2v}{E\sqrt{1-v^2}}b
\\\nnm
&\quad+\frac{E-m_1-m_2}{2}\bigg(a_1+a_2-\frac{m_1-m_2}{E}(a_1-a_2)\bigg),
\end{alignat}
yielding the scattering-angle function $\chi(v,b,\{m,a\})$.

One finds that the result for $\chi=\doe X/\doe b$ is precisely of a form (concerning where various orders begin) able to match (\ref{X1}) plus the $\mathcal{O}(G^2)$ scattering function up to $\mathcal{O}(a^{4(+1)})$ resulting from the test-BH-GOV result (\ref{GOVTBH}) and its parametrized $\mathcal{O}(G^2\sigma^{4})$ corrections arising from (\ref{dmcM4}) --- and plugging such $X_\mr t$ into the EOB map (\ref{EOBmap}) does not change the PN/spin order counting.  Furthermore, matching the $\chi$ from a general Hamiltonian of the above form with a given result for $\chi$ completely determines the quasi-isotropic Hamiltonian coefficients --- up to the freedom in choosing the zeroth-order Hamiltonian $H_0(P^2,\{m\})$.  Two choices are natural: ``real gauge,'' 
\be
H_0^\mr{real}=\sqrt{m_1^2+P^2}+\sqrt{m_2^2+P^2},
\ee
corresponding to $|P|=m_1m_2\gamma v/E$ as $R\to\infty$, \mbox{or ``quasi-isotropic EOB gauge''} \cite{Vines:2018gqi,Buonanno:1998gg},
\be
H^\mr{EOB}_0=\sqrt{m_2^2+2\sqrt{m_1^2m_2^2+P^2(m_1+m_2)^2}+m_1^2},
\ee
corresponding to $|P|=m_1m_2\gamma v/(m_1+m_2)$ as $R\to\infty$.  Once such a choice is made, the scattering-angle coefficients determine all of the quasi-isotropic Hamiltonian coefficients $H_{n}^{\ms a_1\ldots\ms a_\ell}(P^2,\{m\})$ in (\ref{Ham}). This holds even if the actual scattering-angle coefficients have a more complicated mass dependence than is conjectured by the $\mathcal{O}(G^2)$ effective-one-body aligned-spin scattering-angle map (\ref{EOBmap}).

\subsection{From the scattering angle to gauge invariants for circular orbits through subleading post-Newonian orders through fifth order in spin}\label{sec:circ}

Having determined the aligned-spin Hamiltonian coefficients from the scattering-angle coefficients, the former encode the dynamics of both a scattering process and a bound orbit. The Hamiltonian $H(R,P_R,L)$ can then be applied to circular orbits, for which the equations of motion (\ref{Hampolar}) become
\begin{alignat}{3}\label{Hampolar}
\begin{aligned}
0=\dot{P}_R&=-\dfrac{\doe}{\doe R}H(R,0,L),
\\
\Omega=\dot\phi&=\dfrac{\doe}{\doe L}H(R,0,L),
\end{aligned}
\end{alignat}
with $\dot R=0=P_R$ and $\dot L=0=\dot\Omega$. Along with $E=H(R,0,L)$, these can be solved (eliminating $R$) for the energy $E(\Omega)$ and canonical orbital angular momentum $L(\Omega)$ as functions of the orbital frequency $\Omega$.  The results $E(\Omega,\{m,S\})$ and $L(\Omega,\{m,S\})$ are gauge-invariant functions, recalling that the invariant total angular momentum is $J=L+S_1+S_2$.

Under the conjectures ($\dot=$) of the second paragraph of the preamble to Sec.~\ref{sec:genericorbits}, one finds that the combination of interest (for reasons below) $\frak M:=E-\Omega L$ is given, up to subleading post-Newtonian orders through fourth and partial fifth orders in spins, as follows. Let us define the total mass $M$, the reduced mass $\mu$, the symmetric mass ratio $\nu$ and the antisymmetric mass ratio $\delta_m$,
\begin{alignat}{3}
\begin{aligned}
M  = m_1+m_2,
\quad
\mu = \frac{m_1m_2}{M},
\\
\nu =  \frac{\mu}{M},
\quad
\delta_m =  \frac{m_1-m_2}{M},
\end{aligned}
\end{alignat}
the dimensionless (anti)symmetric spin combinations,
\begin{alignat}{3}
\begin{aligned}
\hat a_\pm&=\frac{a_1\pm a_2}{M} =\frac{S_1/m_1\pm S_2/m_2}{m_1+m_2},\phantom{\Bigg|}
\end{aligned}
\end{alignat}
and the PN order counting parameter
\be\label{xPN}
x=(GM\Omega)^{2/3}\sim v^2\sim\frac{GM}{R}.
\ee
We can then express $\mathfrak{M}=E-\Omega L$ as follows, in a PN-spin expansion,
\bse
\begin{alignat}{3}\label{frakMtable}
\begin{aligned}
\frak M =\; M+\mu x\ \big\{ & e^0_1 & \ + & \ e_2^0 x & \ + \ \dots \\
+ & e_1^1\,x^{1.5} & \ + & \ e_2^1\, x^{2.5} & \ + \ \dots \\
+ & e_1^2\,x^{2} & \ + & \ e_2^2\, x^{3} & \ + \ \dots \\
+ & e_1^3\,x^{3.5} & \ + & \ e_2^3\, x^{4.5} & \ + \ \dots \\
+ & e_1^4\,x^{4} & \ + & \ e_2^4\, x^{5} & \ + \ \dots \\
+ & e_1^5\,x^{5.5} & \ + & \ e_2^5\, x^{6.5} & \ + \ \dots \\
& \ \vdots & \ & \ \vdots & \ \ \ \ddots \big\}.
\end{aligned}
\end{alignat}
The conjectures reproduce the known leading-PN coefficients $e_1^\ell$ at $\ell$th order in spins (to all orders in spins), as found by the authors and Steinhoff \cite{Siemonsen:2017yux},
\begin{alignat}{3}
e_1^0&=-\frac{3}{2},
\quad
e_1^1=\frac{7\hat a_++\delta_m\hat a_-}{4},
\quad
e_1^2=-\frac{\hat a_+^2}{2},
\nnm\\
e_1^3&=-\frac{\hat a_+^2}{4}(\hat a_+-\delta_m\hat a_-),
\quad
e_1^{\ell\ge 4}=0,
\end{alignat}
and the known subleading-PN coefficients $e_2^\ell$ \cite{Blanchet:2013haa,Levi:2016ofk,Levi:2015uxa,Levi:2014sba},
\begin{alignat}{3}
8\,e_2^0&=-9-\nu,
\\\nnm
48\,e_2^1&=(99-61\nu)\hat a_++(45-\nu)\delta_m\hat a_-,\phantom{\Big|}
\\\nnm
24\,e_2^2&=(-5+6\nu)\hat a_+^2+-22\hat a_+\delta_m\hat a_-+(1+8\nu)\hat a_-^2,
\end{alignat}
finally producing the conjectural subleading coefficients,
\begin{alignat}{3}
432\,e_2^3&\;\dot=\;(149-666\nu)\hat a_+^3
+3(11-78\nu)\hat a_+^2\delta_m\hat a_-
\\\nnm
&\quad+3(71-296\nu)\hat a_+\hat a_-^2+(5-56\nu)\delta_m\hat a_-^3,\phantom{\Big|}
\\\nnm
32\,e_2^4&\;\dot=\;2\hat a_+^2[2(-2+3\nu)\hat a_+^2+4\hat a_+\delta_m\hat a_-+(-2+8\nu)\hat a_-^2]
\\\nnm
&\quad+(C_\mr{4a}+6C_\mr{4c})(\hat a_+^2+\hat a_-^2)(\hat a_+^2-4\hat a_+\delta_m\hat a_-+\hat a_-^2),\phantom{\Big|}
\end{alignat}
\ese
and the conjectural partial $e_2^5$, modulo $O(G^2a_1^5)$ and $O(G^2a_2^5)$ corrections not considered here, arising from adding $-\frac{1}{5!}{}^{*\!}R_{\mu\rho_1\nu\rho_2;\rho_3\rho_4\rho_5}\sigma^{\rho_1}\ldots\sigma^{\rho_5}$ to (\ref{GOVmcM})$/m^2$ and (\ref{dmcM4}), with no extra $\mc O(R^2\sigma^5)$ terms,
\begin{alignat}{3}\label{e25}
e_2^5&\;\dot=\;\frac{1}{1280}\Big(
({-}347+720\nu)\hat a_+^5
+5(123-176\nu)\hat a_+^4\delta_m\hat a_-
\nnm\\\nnm
&\qquad
+30({-}17+32\nu)\hat a_+^3\hat a_-^2
+10(35-32\nu)\hat a_+^2\delta_m\hat a_-^3
\\\nnm
&\qquad
-135\hat a_+\hat a_-^4
+27\delta_m\hat a_-^5\Big)
\nnm\\\nnm
&\quad+\frac{1}{32}C_\mr{4a}\Big(
\hat a_+^5
+3\hat a_+^4\delta_m\hat a_-
\\\nnm
&\qquad
+3({-}11+8\nu)\hat a_+^3\hat a_-^2
+38({-}35+32\nu)\hat a_+^2\delta_m\hat a_-^3
\\\nnm
&\qquad
+({-}27+16\nu)\hat a_+\hat a_-^4
+7\delta_m\hat a_-^5
\Big)
\\\nnm
&\quad+\frac{3}{8}C_\mr{4c}\Big(
\hat a_+^5
-6\hat a_+^4\delta_m\hat a_-
+2(7-4\nu)\hat a_+^3\hat a_-^2
\\\nnm
&\qquad
-16\hat a_+^2\delta_m\hat a_-^3
+(9-8\nu)\hat a_+\hat a_-^4
-2\hat a_+\delta_m\hat a_-^5
\Big)
\\\nnm
&\quad+\frac{9}{32}C_\mr{4e}\Big(\hat a_-^2-\hat a_+^2\Big)\Big(\hat a_+^3-3\hat a_+^2\delta_m\hat a_-
\\\nnm
&\qquad
+3\hat a_+\hat a_-^2-\delta_m\hat a_-^3\Big)
\\\nnm
&\quad+\mc O(a_1^5)+\mc O(a_2^5).
\end{alignat}



\section{The first law of aligned-spin circular-orbit binary mechanics}\label{sec:firstlaw}

According to the first law(s) of spinning binary (conservative) mechanics, as developed by Le Tiec and collaborators e.g.\ in \cite{Blanchet:2012at,Fujita:2016igj}, for a circular orbit, ``on shell,'' we should have 
\be
\mr d E=\Omega\,\mr d L+\sum_\ms a\Big(z_\ms a\,\mr dm_\ms a+\Omega_\ms a\,\mr d S_\ms a\Big),
\ee
where
\begin{align}
z_\ms a=\frac{\doe H}{\doe m_\ms a}, && \Omega_\ms a=\frac{\doe H}{\doe S_\ms a}
\end{align}
are the redshifts and precession frequencies, respectively, recalling $\Omega=\partial H/\partial L$. It follows that the redshifts are given by
$$
z_\ms a(\Omega,\{m,S\})=\left(\frac{\doe E}{\doe m_\ms a}\right)_{\Omega,\{m,S\}}-\Omega\left(\frac{\doe L}{\doe m_\ms a}\right)_{\Omega,\{m,S\}},
$$
and the precession frequencies are given by
$$
\Omega_\ms a(\Omega,\{m,S\})=\left(\frac{\doe E}{\doe S_\ms a}\right)_{\Omega,\{m,S\}}-\Omega\left(\frac{\doe L}{\doe S_\ms a}\right)_{\Omega,\{m,S\}}.
$$
Equivalently, in terms of the free-energy-like combination presented in the previous section,
$$
\frak M(\Omega,\{m,S\})=E(\Omega,\{m,S\})-\Omega L(\Omega,\{m,S\}),
$$
or $\frak M=E-\Omega L$, the redshifts and the precession frequencies are found from
\begin{align}
z_\ms a=\frac{\doe\frak M}{\doe m_\ms a}, & & \Omega_\ms a=\frac{\doe\frak M}{\doe S_\ms a},
\end{align}
at fixed $\Omega$ (and fixed the other three of $\{m,S\}=\{m_1,m_2,S_1,S_2\}$).
According to the first law, these are the physical redshifts and precession frequencies which would be computed from an appropriately regularized/renormalized spacetime metric.

\subsection*{Redshift and precession frequency through linear deviations from the test-body limit}

From our conjectural $\frak M(\Omega)$ for arbitrary masses and spins from Sec.~\ref{sec:circ}, we can evaluate the redshift $z_1$ and the precession frequency $\Omega_1$ for the smaller body with mass $m_1$ and spin $S_1$, and then consider up to linear deviations from the test-body limit $m_1\to 0$, working to first order in the small (asymmetric) mass ratio
\be
q=\frac{m_1}{m_2},
\ee 
while working to zeroth (or leading) order in the small body's spin $S_1$.  Defining a new PN parameter in terms of the larger mass and the orbital frequency $\Omega$ [cf.\ (\ref{xPN})],
\be
y=(Gm_2\Omega)^{2/3}=\left(\frac{m_2}{M}\right)^{2/3}x,
\ee
and the dimensionless spin of the larger mass,
\be
\hat a=\frac{a_2}{m_2},
\ee
we find the inverse redshift to be
\be
\frac{1}{z_1}=\frac{1}{z_{(0)}}+q\,\delta U+\mc O(q^2),
\ee
where
\be
z_{(0)}=\sqrt{(1- \hat ay^{3/2})\big[1+ \hat ay^{3/2}-3y(1- \hat ay^{3/2})^{1/3}\big]}
\ee
is the exact redshift for circular geodesics in the equatorial plane of the Kerr spacetime, recovered up to the same orders as the following first-order self-force correction:
\begin{alignat}{3}\label{dUconj}
y^{-1}\delta U&=u_1^0
\\\nnm
&\quad+u_2^0y+u_1^1\hat ay^{1.5}+u_1^2\hat a^2y^2
\\\nnm
&\quad\ldots+u_2^1\hat ay^{2.5}+u_2^2\hat a^2y^3+u_1^3\hat a^3y^{3.5}+u_1^4\hat a^4y^4
\\\nnm
&\quad\ldots+u_2^3\hat a^3y^{4.5}+u_2^4\hat a^4y^5+u_1^5\hat a^5y^{5.5}
\\\nnm
&\quad\ldots+\mc O(y^{6}),
\end{alignat}
with coefficients
\begin{alignat}{5}
\begin{aligned}
&u_1^0=-1, \quad u_1^1=\frac{7}{3}, \quad u_1^2=-1,
\\
&u_2^0=-2, \quad u_2^1=\frac{46}{3}, \quad u_2^2=-\frac{86}{9},
\end{aligned}
\end{alignat}
and
\begin{alignat}{3}\label{dUcoeffs}
\begin{aligned}
&u_1^3=1, \quad u_2^3\;\dot=\;\frac{1526}{81},
\\
&u_1^4=0, \quad u_2^4\;\dot=\;{-}(2+C_\mr{4a}+6C_\mr{4c}),
\\
&u_1^5=0.
\end{aligned}
\end{alignat}
For the precession frequency of the small body, we find
\be
\Omega_1=\Omega\Big(\psi_{(0)}+q\,\delta\psi+\mc O(q^2)\Big),
\ee
where 
\be
\psi_{(0)}=1-\sqrt{\frac{1+\hat ay^{3/2}-3y(1-\hat ay^{3/2})^{1/3}}{1-\hat ay^{3/2}}}
\ee
gives the unperturbed-Kerr result, and the first-order self-force correction is
\begin{alignat}{3}\label{dpsi}
\delta\psi&=0 y+\hat ay^{1.5}
\\\nnm
&\quad\ldots+y^2+0 \hat ay^{2.5}+0\hat a^2y^3+0\hat a^3y^{3.5}
\\\nnm
&\quad\ldots-3\hat a^2y^4+\psi_2^3\hat a^3y^{4.5}+0\hat a^4y^5+0\hat a^5y^{5.5}
\\\nnm
&\quad\ldots+\psi_2^4\hat a^4y^6+\mc O(y^{6.5}),
\end{alignat}
with coefficients
\begin{alignat}{3}\label{psicoeffs}
\psi_2^3&\;\dot=\;1,
\qquad\quad
\psi_2^4&\;\dot=\;{-}3(C_\mr{4a}+3C_\mr{4e}).
\end{alignat}

\section{Self-force results from the Teukolsky equation}\label{sec:1SF}

Kavanagh, Ottewill, and Wardell \cite{Kavanagh:2016idg} have computed the linear-in-mass-ratio corrections $\delta U$ to the Kerr-geodesic result $z_{(0)}(\Omega)$ for Detweiler's \cite{Detweiler:2008ft} redshift invariant for circular equatorial orbits in a perturbed Kerr background,
via methods of solving the Teukolsky equation \cite{Teukolsky:1972my} due to Mano, Suzuki and Takasugi \cite{Mano:1996vt,Sasaki:2003xr}.  For a Kerr background with mass $m$ and spin $S=m^2\hat a$, in terms of the orbital frequency $\Omega$ with $y=(Gm\Omega)^{2/3}$, they find
\begin{alignat}{17}\label{dU}
\frac{\delta U}{y}&=-1
\\\nnm
&\quad-2y
\\\nnm
&&&+\tfrac{7}{3}\hat a y^{1.5}
\\\nnm
&\quad-5y^2&&&&-\hat a^2y^2
\\\nnm
&&&+\tfrac{46}{3}\hat a y^{2.5}
\\\nnm
&\quad\;\;\ldots&&&&-\tfrac{86}{9}\hat a^2y^3
\\\nnm
&&&+77\hat ay^{3.5}&&&& +\hat a^3y^{3.5}
\\\nnm
&\quad\;\;\ldots&&&&-\tfrac{577}{9}\hat a^2y^4&&&&+0\hat a^4y^4
\\\nnm
&&&\quad\;\;\,\ldots&&&& \boxed{+\tfrac{1526}{81}\hat a^3y^{4.5}}
\\\nnm
&\quad\;\;\ldots&&&&\quad\;\;\;\ldots&&&&\boxed{-2\hat a^4y^5}
\\\nnm
&\quad+\mc O(y&&{}^{5.5}).
\end{alignat}
In the rearrangement of terms $\widetilde{(n)}$ as suggested by the colors of Figure 1 in \cite{Vines:2016qwa}, coinciding with rows in (\ref{dUconj}) and diagonals in (\ref{dU}),
the purple, blue and green terms are
\begin{alignat}{3}
\frac{1}{y}\delta U^{\widetilde{(0)}}&=-1,
\\\nnm
\frac{1}{y}\delta U^{\widetilde{(1)}}&=-2y+\tfrac{7}{3}\hat ay^{1.5}-\hat a^2y^2,
\\\nnm
\frac{1}{y}\delta U^{\widetilde{(2)}}&=-5y^2
+\tfrac{46}{3}\hat a y^{2.5}
-\tfrac{86}{9}\hat a^2y^3
+\hat a^3y^{3.5}
+0\,\hat a^4y^4,
\end{alignat}
and the yellow terms are
\begin{alignat}{3}\label{dUcoeffGSF}
\frac{1}{y}\delta U^{\widetilde{(3)}}&=\Big({-}\tfrac{121}{3}+\tfrac{123}{16}\zeta_2\Big)y^3+77\hat ay^{3.5}
-\tfrac{577}{9}\hat a^2y^4
\\\nnm
&\quad\boxed{+\tfrac{1526}{81}\hat a^3y^{4.5}
-2\hat a^4y^5}+0\,\hat a^5y^{5.5}+0\,\hat a^6y^6,
\end{alignat}
according to \cite{Kavanagh:2016idg}.

Bini, Damour, Geralico, Kavanagh, and van de Meent \cite{Bini:2018ylh} have computed the first-order self-force correction $\delta\psi$ to the precession frequency for circular equatorial Kerr geodesics to be given by (\ref{dpsi}) with
\begin{align}
\psi_2^3=1, \qquad\qquad\qquad \psi_2^4=0.
\label{psicoeffGSF}
\end{align}

\section{Discussion}\label{sec:conclude}

We have given in Sec.~\ref{sec:firstlaw} the redshift and precession-frequency functions, through linear order in the mass ratio and zeroth order in the smaller body's spin, resulting from (i) the assumption of the existence of a local-in-time canonical Hamiltonian, (ii) the conjecture that the mapping (\ref{EOBmap}) holds with $\ell=4$, through fourth order in spins, producing the arbitrary-mass-ratio binary-BH scattering function from its test-BH limit, and (iii) the conjecture that the test-BH-in-Kerr scattering function must be of the form constructed in Sec.~\ref{sec:TBH}, unique up to cubic order in spins, and parametrized by the effective Wilson coefficients $C_\mr{4a}$, $C_\mr{4c}$ and $C_\mr{4e}$ at order $\mc O(G^2a^4)$.  Recall from \eqref{WCeq1} that the $\mc O(G^2a^4)$ contribution to the scattering-angle function will be finite in the ultrarelativistic limit if
$$
C_{4a}+2C_{4c}+C_{4e}\;\ddot=\;0.
$$
Now we compare the results given in \eqref{dUcoeffs} and \eqref{psicoeffs} to the results obtained for a small body in a circular equatorial orbit in the equatorial plane of a Kerr black hole spacetime \cite{Kavanagh:2016idg,Bini:2018ylh}, given in \eqref{psicoeffGSF} and \eqref{dUcoeffGSF}.  Firstly, at $\mc O(G^2a^3)$, there is a perfect match.  At $\mc O(G^2a^4)$, we find two new constraints on the effective Wilson coefficients:
\begin{align}
\begin{aligned}
C_{4a}+6C_{4c} \;\dot=\;0, \qquad C_{4a}+3C_{4e}\; \dot=\;0.
\end{aligned}
\end{align}
We see that the three equations above have a unique solution,
\begin{align}
C_{4a}\;\ddot=\;C_{4c}\;\ddot=\;C_{4e}\;\ddot=\;0,
\label{punchline}
\end{align}
which coincides with the values corresponding to the result (\ref{GOVTBH}) from GOV \cite{Guevara:2018wpp}.  This confirms the GOV result through linear order in the mass ratio.  Further constraints leading to an overdetermined system of equations could be obtained from new first-order self-force results for eccentric orbits in the equatorial plane of a Kerr spacetime.  Clearly, these exercises can also be continued at $\mc O(G^2a^{>4})$.

These results provide new spin-dependent parts of the conservative dynamics for arbitrary-mass-ratio aligned-spin binary black holes at the fourth-and-a-half and fifth post-Newtonian orders, in particular the subleading post-Newtonian terms at third and fourth orders in the black holes' spins.


\begin{acknowledgements}
We thank Andrea Antonelli, Zvi Bern, Alessandra Buonanno, Eanna Flanagan, Alfredo Guevara, Chris Kavanagh, Mohammed Khalil, Maarten van de Meent, Alexander Ochirov, Jan Steinhoff for helpful discussions and comments on earlier versions of this draft. N.S. was financially supported by the German Academic Scholarship Foundation during the completion of this work. This research was supported in part by Perimeter Institute for Theoretical Physics.  Research at Perimeter Institute is supported by the Government of Canada through the Department of Innovation, Science and Economic Development Canada and by the Province of Ontario through the Ministry of Research, Innovation and Science.
\end{acknowledgements}

\bibliography{TBH}


\end{document}